\documentclass[prc,preprint,superscriptaddress,amsmath,amssymb]{revtex4}
\usepackage{graphicx}
\usepackage{bm}

\newcommand{\bra}[1]{\langle {#1} |}
\newcommand{\ket}[1]{| {#1} \rangle}
\newcommand{\inproduct}[2]{\langle #1 | #2 \rangle}
\newcommand{\vecr}{{\mathbf r}}

\begin{document}

\title{
Finite amplitude method for the RPA solution\\
}

\author{Takashi Nakatsukasa}
\affiliation{Center for Computational Sciences
University of Tsukuba, Tsukuba 305-8571, Japan}
\affiliation{Institute of Physics, University of Tsukuba, 
Tsukuba 305-8571, Japan}
\author{Tsunenori Inakura}
\affiliation{Institute of Physics, University of Tsukuba, 
Tsukuba 305-8571, Japan}
\author{Kazuhiro Yabana}
\affiliation{Center for Computational Sciences
University of Tsukuba, Tsukuba 305-8571, Japan}
\affiliation{Institute of Physics, University of Tsukuba, 
Tsukuba 305-8571, Japan}

\date{\today}

\begin{abstract}
We propose a practical method to solve
the random-phase approximation (RPA)
in the self-consistent Hartree-Fock (HF) and density-functional theory.
The method is based on numerical evaluation of the residual interactions
utilizing finite amplitude of single-particle wave functions.
The method only requires calculations of the single-particle
Hamiltonian constructed with independent bra and ket states.
Using the present method,
the RPA calculation becomes possible with a little extension
of a numerical code of the static HF calculation.
We demonstrate usefulness and accuracy of the present method 
performing test calculations for isoscalar responses
in deformed $^{20}$Ne.
\end{abstract}

\maketitle


\section{Introduction}
\label{sec: introduction}

The mean-field theory with a density-dependent effective interaction has
been an essential tool to understand nuclei.
Thanks to the high performance computing, 
it is now becoming the most promising tool for quantitative description of
nuclear structure in medium-to-heavy nuclei \cite{BHR03,LPT03}.
The nuclear self-consistent mean-field theories are analogous to to the
density-functional theory in condensed matter.
A current major goal is constructing a universal energy-density functional,
which is able to describe ground and excited states in nuclei and
nuclear matter.
This is also urgently needed for predicting and interpreting new data from
the next generation of radioactive beam facilities.

In order to describe dynamical properties in nuclear response to external
fields, the random-phase approximation (RPA) is a
leading theory applicable to both low-lying states and giant resonances
\cite{RS80}.
The RPA is a microscopic theory which can be obtained by linearizing the
time-dependent Hartree-Fock (TDHF) equation, or equivalently,
the time-dependent Kohn-Sham equation in the density-functional theory.
The linearization produces a self-consistent residual interaction,
$v = \delta^2 E[\rho]/\delta\rho^2$,
where $E$ and $\rho$ are the energy-density functional and
the one-body density,
respectively (Sec.~\ref{sec: linear_response_theory}).
The standard solution of the RPA is based on the matrix formulation
of the RPA equation, which involves a large number of particle-hole
matrix elements of the residual interaction, $v_{ph',hp'}$ and $v_{pp',hh'}$.
Since the realistic nuclear energy functional is rather complicated,
it is very tedious and difficult to
calculate all the necessary matrix elements.
It is, therefore, the purpose of the present paper to present an alternative
method of solving the RPA equations, in which we deal with only
the single-particle Hamiltonian, $h[\rho]$.

Although there are numerous works on the HF-plus-RPA calculations,
because of the complexity of the residual interactions,
it has been common in practice
to neglect some parts of the residual interactions.
The RPA calculations with full self-consistency are becoming a current
trend in nuclear structure studies, however, they are essentially
only for spherical nuclei at present \cite{Ter05,Gia03,VPRL01,PRNV03}.
The applications to deformed nuclei are very few, but have been done
for the Skyrme energy functional using the three-dimensional
mesh-space representation \cite{NY02-P2,IH03,Ina04,NY05}.
See Sec.~I in Ref.~\cite{Ter05} for a current status of these studies.

The basic idea of the present method is analogous to linear-response
calculations in a time-dependent manner (real-time method)
\cite{NY02-P2,NY05,UO05}.
In the real-time method, the time evolution of a TDHF state involves only
the action of the HF Hamiltonian, $h[\rho(t)]$,
onto single-particle orbitals, $\ket{\psi_i(t)}$ ($i=1,\cdots,A$).
Although the real-time method is very efficient for obtaining nuclear
response in a wide energy range,
its numerical instability caused by zero modes was a problem for
the linear-response calculations \cite{NY05}.
Zero-energy modes related to symmetry breaking in the HF state
are easily excited, which often prevents the calculation
of the time evolution for a long period.
Therefore, it is desirable to develop a corresponding method
in the frequency (energy) representation.

This paper is organized as follows.
A new approach to the solution of the linear response equation,
``finite amplitude method'',
is presented in Sec.~\ref{sec: linear_response_theory}.
In Sec.~\ref{sec: numerical_applications}, using the
Bonche-Koonin-Negele (BKN) interaction \cite{BKN76},
we check the accuracy of solutions obtained with the present method.
We also investigate the zero-energy components in calculated
strength functions.
Then, the conclusion is summarized in Sec.~\ref{sec: conclusion}.

\section{Linear response theory}
\label{sec: linear_response_theory}

The RPA equation is known to be equivalent to the TDHF equation
in the small-amplitude limit.
We recapitulate how the RPA equation is derived from the small-amplitude
TDHF equation, that will help explanation of our basic idea.

\subsection{TDHF and linear response equation}

The HF Hamiltonian, $h[\rho]=\delta E[\rho]/\delta\rho$,
is a functional of one-body density
matrix, $\rho$, which satisfies
the condition, $\rho^2 = \rho$,
that the state is expressed by a single Slater determinant.
The stationary condition is
\begin{equation}
\label{HF}
[h[\rho],\rho] = 0,
\end{equation}
which defines the HF ground state density $\rho=\rho_0$.
Hereafter, the static HF Hamiltonian is simply denoted as $h_0=h[\rho_0]$,
and $\hbar=1$ is used.
When a time-dependent external perturbation is present,
the time evolution of the density, $\rho(t)$, follows
the TDHF equation,
\begin{equation}
\label{TDHF}
i\frac{d}{dt}\rho(t) = \big[\ h[\rho(t)]+V_{\rm ext}(t),\ \rho(t)\ \big] .
\end{equation}
Using this $\rho(t)$,
the expectation value of a one-body operator $F$ is obtained as
$\langle F \rangle={\rm tr} \{F\rho(t)\}$.
Provided that the perturbation is weak,
we may linearize Eq.~(\ref{TDHF}) with respect to 
$V_{\rm ext}(t)$ and $\delta\rho(t)$ defined by
\begin{equation}
\label{linear_approx}
\rho(t)=\rho_0+\delta\rho(t) .
\end{equation}
This leads to a time-dependent linear-response equation
with an external field,
\begin{equation}
\label{LTDHF}
i\frac{d}{dt}\delta\rho(t) = [h_0,\delta\rho(t)]
                       +[V_{\rm ext}(t)+\delta h(t),\rho_0] ,
\end{equation}
where $\delta h(t)$ is a residual field induced by density fluctuations,
\begin{equation}
\label{delta_h_t}
\delta h(t) \equiv \frac{\delta h}{\delta \rho}\cdot \delta\rho(t)
   = \sum_{\mu\nu}
     \left.\frac{\partial h}{\partial \rho_{\mu\nu}}\right|_{\rho=\rho_0}
     \delta\rho_{\mu\nu}(t) .
\end{equation}
It should be noted that $\delta h(t)$ has a linear dependence on
$\delta\rho(t)$.
As we will see in Eq. (\ref{cond_delta_rho}),
if we adopt the natural basis diagonalizing $h_0$,
the summation can be restricted to the particle-hole ($\mu> A, \nu\le A$)
and hole-particle ($\mu\ge A, \nu< A$) components.
Now, we decompose the time-dependent $\delta\rho(t)$ into
those with fixed frequencies:
\begin{equation}
\label{delta_rho_omega}
\delta\rho(t)=\sum_\omega \big\{ \eta \delta\rho(\omega) e^{-i\omega t}
                       +\eta^* \delta\rho^\dagger(\omega) e^{i\omega t} \big\} .
\end{equation}
The external and induced fields are also expressed in the same way.
\begin{eqnarray}
\delta h(t)&=&\sum_\omega \left\{ \eta \delta h(\omega) e^{-i\omega t} 
                  + \eta^* \delta h^\dagger (\omega) e^{i\omega t} \right\} ,\\
\label{Vext}
V_{\rm ext}(t)&=&\sum_\omega \left\{ \eta V_{\rm ext}(\omega) e^{-i\omega t}
                  + \eta^* V_{\rm ext}^\dagger (\omega) e^{i\omega t} \right\} .
\end{eqnarray}
Here, we have introduced a small dimensionless parameter $\eta$.
$\delta h(\omega)$ may be written as
$\delta h(\omega)=\delta h/\delta\rho \cdot \delta\rho(\omega)$.
Note that the transition density, the external field,
and the induced field in the $\omega$-representation,
$\delta\rho(\omega)$, $V_{\rm ext}(\omega)$, and $\delta h(\omega)$,
are not necessarily hermitian.
Substituting these into the linearized TDHF equation (\ref{LTDHF}),
we obtain the linear-response equation in the frequency representation, 
\begin{equation}
\label{LRE}
\omega\ \delta\rho(\omega) = [h_0,\delta\rho(\omega)]
                       +[V_{\rm ext}(\omega)+\delta h(\omega),\rho_0] .
\end{equation}
This is the equation we want to solve in this paper.

When the frequency $\omega$
is equal to an RPA eigenfrequency $\omega_n$,
there is a non-zero solution, $\delta\rho_n$, of Eq.~(\ref{LRE})
with $V_{\rm ext}=0$.
These are called normal modes and are orthogonal to each other.
The normalization is given by
\begin{equation}
\label{RPA_orthonormalization}
\text{tr}\{ [ \delta\rho_n^\dagger, \delta\rho_{n'} ] \rho_0 \} =
\bra{\Phi_0} [ \delta\rho_n^\dagger, \delta\rho_{n'} ] \ket{\Phi_0}
= \delta_{nn'} ,
\end{equation}
where $\ket{\Phi_0}$ indicates the HF ground state.
From Eq.~(\ref{RPA_orthonormalization}), it is obvious that,
in order to normalize the transition density $\delta\rho_n$,
it must be non-hermitian.
When $\omega=\omega_n$,
the nucleus is truly excited by $V_{\rm ext}(\omega)$,
and we cannot determine the magnitude of $\delta\rho(\omega_n)$
because $\delta\rho(t)$ increases in time.
If $\delta\rho(\omega_n)$ is a solution of Eq.~(\ref{LRE}),
$\delta\rho(\omega_n) + c \delta\rho_n$ with an arbitrary constant $c$
is a solution too.

So far, the linear-response equation has been expressed in terms of
the one-body density operators.
The density-matrix formulation is simple and easy to manipulate, however,
in practical calculations,
it is convenient to introduce single-particle (Kohn-Sham) orbitals.
For systems with $A$ particles,
the TDHF describes the one-body density using $A$ single-particle
orbitals, $\ket{\psi_i(t)}$,
\begin{equation}
\label{TDHF_orbitals}
\rho(t)=\sum_{i=1}^A \ket{\psi_i(t)}\bra{\psi_i(t)} ,
\quad\quad\quad
\rho_0=\sum_{i=1}^A \ket{\phi_i}\bra{\phi_i} .
\end{equation}
It is an advantage of the TDHF, that the time evolution is described
by occupied orbitals only, $\{ \ket{\psi_i} \}$ with $i=1,\cdots,A$.
The static orbitals are normally chosen as eigenstates of the HF Hamiltonian,
\begin{equation}
h_0 \ket{\phi_\mu} = \epsilon_\mu \ket{\phi_\mu} ,
\end{equation}
which can be divided into two categories;
occupied (hole) orbitals, $\{\phi_i\}$ ($i=1,\cdots,A$),
for which we use indexes $i,j,\cdots$, and
unoccupied (particle) orbitals, $\{\phi_m\}$ ($m=A+1,\cdots$),
for which we use indexes $m,n,\cdots$.
In the linear approximation, we have
\begin{equation}
\label{delta_rho}
\delta\rho(t)=\sum_i \big\{
    \ket{\phi_i}\bra{\delta\psi_i(t)}
   +\ket{\delta\psi_i(t)}\bra{\phi_i} \big\} ,
\end{equation}
where $\ket{\psi_i(t)}=(\ket{\phi_i}+\ket{\delta\psi_i(t)})e^{-i\epsilon_i t}$
and it is linearized with respect to $\ket{\delta\psi_i(t)}$.
The condition, $\rho(t)^2 = \rho(t)$,
leads to
\begin{eqnarray}
\label{cond_delta_rho}
&&\delta\rho_{ij}=\delta\rho_{mn}=0 , \quad\quad i,j \leq A, \quad m,n >A ,\\
&&\inproduct{\phi_j}{\delta\psi_i}+\inproduct{\delta\psi_j}{\phi_i}=0 .
\end{eqnarray}
The second equation is nothing but the 
orthonormalization condition for single-particle orbitals,
$\{\ket{\psi_i(t)}\}$ ($i=1,\cdots,A$).

Transforming $\delta\rho(t)$ into $\delta\rho(\omega)$ in
Eq. (\ref{delta_rho_omega}),
we must make ket and bra states independent,
because $\delta\rho(\omega)$ is not hermitian.
This is related to the fact that the RPA equation is described by
forward and backward amplitudes, $X(\omega)$ and $Y(\omega)$.
\begin{equation}
\label{delta_rho_XY}
\delta\rho(\omega) = \sum_i \big\{ \ket{X_i(\omega)}\bra{\phi_i}
                       + \ket{\phi_i}\bra{Y_i(\omega)} \big\} .
\end{equation}
This is equivalent to the Fourier decomposition of the
time-dependent single-particle orbitals,
\begin{equation}
\ket{\delta\psi_i(t)}=\sum_\omega \big\{ \eta \ket{X_i(\omega)} e^{-i\omega t}
                       + \eta^* \ket{Y_i(\omega)} e^{i\omega t} \big\} .
\end{equation}
Since only the particle-hole matrix elements of $\delta\rho(\omega)$ are
non-zero,  seen in Eq. (\ref{cond_delta_rho}),
we can assume that the amplitudes, $\ket{X_i(\omega)}$ and $\ket{Y_i(\omega)}$,
can be expanded in the particle orbitals only;
\begin{equation}
\ket{X_i(\omega)}=\sum_{m>A} \ket{\phi_m} X_{mi}(\omega) ,
\quad\quad
\ket{Y_i(\omega)}=\sum_{m>A} \ket{\phi_m} Y_{mi}^*(\omega) .
\end{equation}
If we take particle-hole and hole-particle matrix elements of Eq.~(\ref{LRE}),
we can derive the well-known RPA equation in the matrix form;
\begin{equation}
\label{RPA}
\left\{
\begin{pmatrix}
A   & B \\
B^* & A^*
\end{pmatrix}
- \omega
\begin{pmatrix}
1 & 0 \\
0 & -1
\end{pmatrix}
\right\}
\begin{pmatrix}
X_{nj}(\omega) \\
Y_{nj}(\omega)
\end{pmatrix}
= -
\begin{pmatrix}
f(\omega) \\
g(\omega)
\end{pmatrix} .
\end{equation}
Here, the matrices, $A$ and $B$, and the vectors, $f$ and $g$,
are defined by
\begin{eqnarray}
A_{mi,nj}&\equiv& (\epsilon_m-\epsilon_i)\delta_{mn}\delta_{ij}
+ \bra{\phi_m}
\left.\frac{\partial h}{\partial\rho_{nj}}\right|_{\rho=\rho_0}\ket{\phi_i} ,\\
B_{mi,nj}&\equiv& \bra{\phi_m}
\left.\frac{\partial h}{\partial\rho_{jn}}\right|_{\rho=\rho_0}\ket{\phi_i}, \\
f_{mi}(\omega)&\equiv& \bra{\phi_m} V_{\rm ext}(\omega) \ket{\phi_i} ,\quad
g_{mi}(\omega)\equiv \bra{\phi_i} V_{\rm ext}(\omega) \ket{\phi_m} .
\end{eqnarray}
For the normal modes that are homogeneous solutions of Eq. (\ref{RPA}),
the orthogonality and normalization are expressed as
\begin{equation}
\sum_{m,i} \left\{ X_{mi}^{(n)*} X_{mi}^{(n')}
    - Y_{mi}^{(n)*} Y_{mi}^{(n')} \right\} = \delta_{nn'} ,
\end{equation}
which is equivalent to Eq. (\ref{RPA_orthonormalization}).
This is a standard matrix formulation of the RPA equation.
In practical applications, the most tedious part is calculation of
matrix elements of the residual interactions in $A_{mi,nj}$ and $B_{mi,nj}$.
In Ref.~\cite{Muta02}, a numerical method to solve the RPA equation
in the coordinate space is proposed, and the similar approaches are used
in realistic applications using the Skyrme interaction \cite{IH03,Ina04}.
In those works, one does not need to calculate the particle orbitals,
however, the residual interaction must be evaluated in the coordinate-space
representation.
In Sec.~\ref{sec: FAM},
we propose an alternative, even simpler approach to
a solution of the linear-response equation (\ref{LRE}).
The method does not require explicit evaluation of the residual
interaction, $\delta h/\delta\rho$.

\subsection{Finite amplitude method}
\label{sec: FAM}

Multiplying both sides of Eq.~(\ref{LRE}) with a ket of hole states
$\ket{\phi_i}$,
we have
\begin{equation}
\label{RPA1}
\omega \ket{X_i(\omega)} = (h_0-\epsilon_i)\ket{X_i(\omega)}
+ \hat{Q} \left\{ V_{\rm ext}(\omega) + \delta h(\omega)\right\} \ket{\phi_i},
\end{equation}
where $\hat{Q}$ is a projection operator onto the particle space,
$\hat{Q}=1-\sum_j \ket{\phi_j}\bra{\phi_j}$.
Another equation can be derived by multiplying a bra state $\bra{\phi_i}$ with
Eq.~(\ref{LRE}).
\begin{equation}
\label{RPA2}
\omega \bra{Y_i(\omega)} = -\bra{Y_i(\omega)}(h_0-\epsilon_i)
- \bra{\phi_i} \left\{ V_{\rm ext}(\omega) + \delta h(\omega)\right\} \hat{Q}.
\end{equation}
These are formally equivalent to the RPA equation in the matrix form of
Eq.~(\ref{RPA}).

The essential idea of our new numerical approach is as follow:
Equations (\ref{RPA1}) and (\ref{RPA2}) require operations of the
HF Hamiltonian in the ground state, $h_0$,
and the induced fields, $\delta h(\omega)$ and
$\delta h^\dagger(\omega)$.
Since $h_0$ is obtained by the static HF calculation,
a new ingredient for the RPA calculation is the latter two.
The conventional approach is to expand $\delta h(\omega)$
in the linear order as Eq.~(\ref{delta_h_t}),
then to solve the RPA equation in a matrix form.
In this paper, instead of performing the explicit expansion,
we resort to the numerical linearization.
Now, let us explain how to achieve it.

The time-dependent self-consistent Hamiltonian,
$h(t)$,
is a functional of one-body density that is represented by occupied
$A$ single-particle states; $h[\rho(t)]=h[\psi(t)]$.
In the linear approximation,
\begin{equation}
h[\rho_0+\delta\rho(t)]=h[\phi+\delta\psi(t)]=h_0+\delta h(t) ,
\end{equation}
the induced field can be written as
$\delta h(t)=h[\phi+\delta\psi(t)]-h_0$.
In the frequency representation,
the story becomes slightly more complicated, because
$\delta h(\omega)$ and $\delta h^\dagger (\omega)$, are no longer hermitian.
In this case, we should regard $h[\rho]$ 
as a functional of $2A$ single-particle states (independent bra and ket),
$\bra{\psi'_i}$ and $\ket{\psi_i}$, $i=1,\cdots,A$.
We denote it as $h\big[\bra{\psi'},\ket{\psi}\big]$.
Using Eq. (\ref{delta_rho_XY}),
we may write the non-hermitian density as
\begin{eqnarray}
\rho_0+\eta \delta\rho(\omega)
&=&\sum_i \left\{
\ket{\phi_i}\bra{\phi_i} + \eta \ket{X_i(\omega)}\bra{\phi_i}
+\eta \ket{\phi_i}\bra{Y_i(\omega)}
\right\} \\
&=&\sum_i \left\{\ket{\phi_i}+\eta\ket{X_i(\omega)}\right\}
\left\{\bra{\phi_i}+\eta\bra{Y_i(\omega)}\right\} .
\end{eqnarray}
In the last equation, we assume the linear approximation with respect to $\eta$.
The fact that
$\delta h(\omega)$ is proportional to $\delta\rho(\omega)$ and
$\delta h^\dagger(\omega)$ proportional to $\delta\rho^\dagger(\omega)$
leads to
\begin{eqnarray}
h_0+\eta\ \delta h(\omega) &=&
h[\rho_0+\eta\delta\rho(\omega)]=
h\big[\bra{\phi}+\eta\bra{Y(\omega)},\ket{\phi}+\eta\ket{X(\omega)}\big] ,\\
h_0+\eta\ \delta h^\dagger(\omega) &=&
h[\rho_0+\eta\delta\rho^\dagger(\omega)]=
h\big[\bra{\phi}+\eta\bra{X(\omega)},\ket{\phi}+\eta\ket{Y(\omega)}\big] ,
\end{eqnarray}
up to the first order in a small parameter $\eta$.
In other words, the induced fields may be calculated using the
finite difference with respect to $\eta$;
\begin{equation}
\label{delta_h_2}
\delta h(\omega) = \frac{1}{\eta} 
\left(
h\big[\bra{\psi'},\ket{\psi}\big] - h[\bra{\phi},\ket{\phi}\big] \right),
\end{equation}
where $\bra{\psi'_i}=\bra{\phi_i}+\eta \bra{Y_i(\omega)}$ and
$\ket{\psi_i}=\ket{\phi_i}+\eta \ket{X_i(\omega)}$.
Its hermitian conjugate, $\delta h^\dagger(\omega)$,
may be expressed as the same equation
(\ref{delta_h_2}), but with
$\bra{\psi'_i}=\bra{\phi_i}+\eta \bra{X_i(\omega)}$ and
$\ket{\psi_i}=\ket{\phi_i}+\eta \ket{Y_i(\omega)}$.

Using these numerical differentiation,
the r.h.s. of the RPA equations, (\ref{RPA1}) and (\ref{RPA2}),
can be easily calculated by action of the HF Hamiltonian,
$h\big[\bra{\psi'},\ket{\psi}\big]$, on the single-particle orbitals,
$\ket{\phi_i}$.
For the first sight,
Eqs.~(\ref{RPA1}) and (\ref{RPA2}) do not look like linear equations.
However, since $\delta h(\omega)$ linearly depends on 
$\ket{X_i(\omega)}$ and $\bra{Y_i(\omega)}$,
they are inhomogeneous linear equations with respect to
$\ket{X_i(\omega)}$ and $\bra{Y_i(\omega)}$.
It is obvious in a sense that they are equivalent to the matrix form
of Eq.~(\ref{RPA}),
Therefore, we can employ a well-established iterative method
for their solutions.
If the linear equation is described by a hermitian matrix, the conjugate
gradient method (CGM) is one of the most powerful method.
However, in general, we may take the frequency $\omega$ complex,
then, the RPA matrix becomes non-hermitian.
Then, we should use another kind of iterative solver,
for instance, the bi-conjugate gradient method (Bi-CGM).
A typical numerical procedure is as follows:
(i) Fix the frequency $\omega$ that can be complex, and
assume initial vectors ($n=0$),
$\ket{X_i^{(n)}(\omega)}$ and $\bra{Y_i^{(n)}(\omega)}$.
(ii) Update the vectors,
$\ket{X_i^{(n+1)}(\omega)}$ and $\bra{Y_i^{(n+1)}(\omega)}$,
    using the algorithm of an iterative method, such as CGM and Bi-CGM.
(iii) Calculate the residual of Eqs. (\ref{RPA1}) and (\ref{RPA2}).
    If its magnitude is smaller than a given accuracy,
    stop the iteration.
    Otherwise, go back to the step (ii) with $n\rightarrow n+1$.

The most advantageous feature of the present approach
is that it only requires operations of the HF Hamiltonian,
$h\big[\bra{\psi'},\ket{\psi}\big]$.
These are usually included in computational programs of the
static HF calculations.
Only extra effort necessary is to estimate the HF Hamiltonian 
with different bra and ket single-particle states,
$\bra{\psi_i'}$ and $\ket{\psi_i}$.
Therefore, a minor modification of the static HF computer code
will provide a numerical solution of the RPA equations.
Hereafter, we call this numerical approach
``finite amplitude method''.
Apparently, the present method is also applicable to the RPA eigenvalue
problems with a trivial modification.

\subsection{Transition strength in the linear response}

In this subsection, we present how to calculate transition strength
using solutions of Eqs. (\ref{RPA1}) and (\ref{RPA2}).
Assuming that the system is at its ground state $\ket{\Phi_0}$
with energy $E_0=0$ at $t=-\infty$,
and that the external field $V_{\rm ext}(t)$ is
adiabatically switched on
($\omega\rightarrow\omega\pm i\epsilon$ in Eq.~(\ref{Vext})),
the state at time $t$ will be
\begin{equation}
\ket{\Psi(t)}=\ket{\Phi_0}-i\sum_n e^{-iE_n t} \int_{-\infty}^t dt'
e^{iE_n t'} \ket{\Phi_n}\bra{\Phi_n}V_{\rm ext}(t')\ket{\Phi_0} .
\end{equation}
in the first-order approximation with respect to $V_{\rm ext}$.
Here, $\ket{\Phi_n}$ and $E_n$ are the $n$-th excited state and its
excitation energy, respectively.
Especially, if the external field has a fixed frequency $\omega>0$,
$V_{\rm ext}(t)=\eta F e^{-i\omega t} + \eta^* F^\dagger e^{i\omega t}$,
this is written as
\begin{equation}
\ket{\Psi(t)}=\ket{\Phi_0}-i\sum_n \ket{\Phi_n}\left(
\frac{\eta\bra{\Phi_n}F\ket{\Phi_0}}{\omega-E_n+i\epsilon} e^{-i\omega t}
-\frac{\eta^*\bra{\Phi_n}F^\dagger\ket{\Phi_0}}{\omega+E_n-i\epsilon}
  e^{i\omega t} \right) ,
\end{equation}
where $F$ is an arbitrary one-body operator,
Then, the expectation value of $F^\dagger$ at time $t$ is
\begin{eqnarray}
\label{Psi_F_Psi}
\bra{\Psi(t)}F^\dagger\ket{\Psi(t)}
&\equiv&
\bra{\Phi_0}F^\dagger\ket{\Phi_0} + \eta S(F;\omega) e^{-i\omega t} + \cdots,\\
S(F;\omega)&=&\sum_n\left(
\frac{|\bra{\Phi_n}F\ket{\Phi_0}|^2}{\omega-E_n+i\epsilon}
- \frac{|\bra{\Phi_n}F^\dagger\ket{\Phi_0}|^2}{\omega+E_n-i\epsilon}
\right) .
\end{eqnarray}
Taking the limit of $\epsilon\rightarrow 0$, we have the transition strength,
\begin{equation}
\frac{dB(\omega; F)}{d\omega} \equiv
 \sum_n |\bra{\Phi_n}F\ket{\Phi_0}|^2 \delta(\omega-E_n)
 = -\frac{1}{\pi} \mbox{Im} S(F;\omega) .
\end{equation}
Comparing Eq. (\ref{Psi_F_Psi}) with the expectation value
in the TDHF state,
\begin{equation}
\mbox{tr}\big\{F^\dagger\rho(t)\big\}
 =\mbox{tr}\big\{F^\dagger \rho_0\big\}
 +\mbox{tr}\big\{F^\dagger \delta\rho(\omega)\big\} e^{-i\omega t} +\cdots .
\end{equation}
$S(F:\omega)$ in the RPA is written as
\begin{eqnarray}
S_{\rm RPA}(F;\omega) &=& \mbox{tr}\big\{F^\dagger \delta\rho(\omega)\big\}
=i \mbox{tr}\big\{ [ \delta\rho_F^\dagger,\delta\rho(\omega)] \big\} \\
&=&\sum_i \left( \bra{\phi_i}F^\dagger\ket{X_i(\omega)}
+\bra{Y_i(\omega)}F^\dagger\ket{\phi_i} \right) .
\end{eqnarray}
Here, $\delta\rho_F$ is defined by
$\delta\rho_F\equiv i [F,\rho_0]$.

\subsection{Separation of Nambu-Goldstone modes}
\label{sec: NG_mode}

The RPA theory is known to have a property that the zero-energy modes are
exactly decoupled from physical (intrinsic) modes of excitation.
Since the zero modes are associated with the symmetry breaking in the
HF ground state, it is also called ``Nambu-Goldstone modes'' (NG modes).
When $P$ is a hermitian symmetry operator of the Hamiltonian,
$[ H, P ] = 0 $,
then, the transformed ground-state density,
$\tilde{\rho}_0=e^{i\alpha P}\rho_0 e^{-i\alpha P}$,
also satisfies the HF equation (\ref{HF}).
Expanding the equation up to the first order in $\alpha$, we have
\begin{equation}
[ h_0, \delta\tilde{\rho}] + [ \delta \tilde{h},\rho_0] = 0 ,
\end{equation}
where
\begin{equation}
\delta\tilde{\rho} \equiv \tilde{\rho}_0 - \rho_0 = i\alpha [P,\rho_0],
\quad\quad
\delta \tilde{h} \equiv h[\tilde{\rho}_0] - h_0
= \frac{\delta h}{\delta \rho}\cdot\delta\tilde{\rho}.
\end{equation}
This indicates that $\delta\tilde{\rho}$ is an RPA eigenmode corresponding
to $\omega=0$.
$\delta\rho$ generated by the operator $R$ conjugate to $P$
($[R,P]=i$) can be defined in a similar manner.
For instance, the translational symmetry is expressed by
the total momentum as $P$ and the center-of-mass coordinate as $R$.
We denote these transition densities associated with the NG mode as
\begin{eqnarray}
\delta\rho_P&\equiv& i [P,\rho_0]= \frac{1}{\alpha}\delta\tilde{\rho}
 =\sum_i \left(\ket{\bar{P}_i}\bra{\phi_i}+\ket{\phi_i}\bra{\bar{P}_i}\right)
 , \\
\delta\rho_R&\equiv& i [R,\rho_0] 
 =\sum_i \left(\ket{\bar{R}_i}\bra{\phi_i}+\ket{\phi_i}\bra{\bar{R}_i}\right) ,
\end{eqnarray}
where we have defined
$\ket{\bar P_i}\equiv iP\ket{\phi_i}$ and
$\ket{\bar R_i}\equiv iR\ket{\phi_i}$.
Provided that $P$ and $R$ are hermitian, $\delta\rho_P$ and $\delta\rho_Q$
are also hermitian.
Therefore, we cannot normalize them in terms of the normalization condition
of Eq.~(\ref{RPA_orthonormalization}).
These modes are automatically orthogonal to other normal modes with
$\omega\neq 0$.
If we solve the RPA equation fully self-consistently,
the NG modes should be clearly separated from other modes.
However, in practice, we often encounter a mixture
of spurious components in physical excitations.
For instance, the coordinate-space is discretized in mesh to represent
wave functions in Sec.~\ref{sec: numerical_applications},
which violates the exact translational and rotational symmetries.
We also use a smoothing parameter $\Gamma$ to make
the frequency complex, then,
low-lying excited states are embedded in a large tail of the
NG-mode strength
($\delta\rho(\omega)\rightarrow\infty$ for $\omega\rightarrow 0$ ).
Here, we present a prescription to remove the strength associated with the
NG mode.

Let us assume that there is a mixture of NG modes in
a calculated transition density, $\delta\rho_{\rm cal}(\omega)$.
\begin{equation}
\label{delta_rho_phy}
\delta\rho_{\rm cal}(\omega)=\delta\rho_{\rm phy}(\omega)
 +\lambda_P \delta\rho_P+\lambda_R \delta\rho_R ,
\end{equation}
where ``physical'' transition density, $\delta\rho_{\rm phy}(\omega)$,
is free from the NG modes.
Here, we assume there is a single NG mode, for simplicity.
It is straightforward to extend
the present prescription to the one for more than one NG modes.
Since $\delta\rho_{\rm phy}$ should be orthogonal to the NG modes,
$\delta\rho_{\rm phy}$ should satisfy
\begin{equation}
\label{NG_orthogonal_condition}
\bra{\Phi_0} [\delta\rho_P,\delta\rho_{\rm phy}(\omega)] \ket{\Phi_0}=
\bra{\Phi_0} [\delta\rho_R,\delta\rho_{\rm phy}(\omega)] \ket{\Phi_0}= 0.
\end{equation}
Utilizing the canonicity condition, $[R,P]=i$,
the orthogonality condition, Eq. (\ref{NG_orthogonal_condition}),
determines the coefficients, $\lambda_{P(R)}$, as
\begin{eqnarray}
\lambda_P&=&-i\sum_i\left(\inproduct{\bar{R}_i}{X_i(\omega)}
                       -\inproduct{Y_i(\omega)}{\bar{R}_i} \right),\\
\lambda_R&=&i\sum_i\left(\inproduct{\bar{P}_i}{X_i(\omega)}
                       -\inproduct{Y_i(\omega)}{\bar{P}_i} \right).
\end{eqnarray}
Substituting these into Eq. (\ref{delta_rho_phy}),
we may extract $\delta\rho_{\rm phy}(\omega)$
from the ``contaminated'' transition density $\delta\rho_{\rm cal}(\omega)$.

\section{Numerical Applications}
\label{sec: numerical_applications}

\subsection{Coordinate-space representation}

In case of zero-range effective interactions, such as Skyrme interactions,
the HF Hamiltonian, $h(\vecr)=h[\rho(\vecr)]$,
is a functional of local one-body densities.
Then, it is convenient to adopt the coordinate-space representation.
In the followings, we assume $\vecr$ involves the spin and isospin
indexes, if necessary.
The RPA equations, (\ref{RPA1}) and (\ref{RPA2}), for
a complex frequency $\omega$ can be written in the $\vecr$-representation as
\begin{eqnarray}
\label{RPA1_2}
\big( h_0(\vecr)-\epsilon_i-\omega\big)X_i(\vecr,\omega)
+ \delta h(\vecr,\omega) \phi_i(\vecr)
&=&- V_{\rm ext}(\vecr,\omega) \phi_i(\vecr) ,\\
\label{RPA2_2}
\left\{
\big(h_0(\vecr)-\epsilon_i+\omega^*\big)Y_i(\vecr,\omega)
+ \delta h^\dagger(\vecr,\omega) \phi_i(\vecr) \right\}^*
&=&-\left\{V_{\rm ext}^\dagger(\vecr,\omega) \phi_i(\vecr) \right\}^* .
\end{eqnarray}
Here, for simplicity, we omit the projection operator, $\hat{Q}$,
on both sides of these equations.
In the finite amplitude method,
the operation of $\delta h(\vecr,\omega)$ is calculated by
\begin{equation}
\label{delta_h_3}
\delta h(\vecr,\omega)\phi_i(\vecr) = \frac{1}{\eta} 
\left(
h\big[\psi'^{*},\psi\big](\vecr)\phi_i(\vecr)
    - \epsilon_i\phi_i(\vecr) \right),
\end{equation}
with $\psi'^*_i(\vecr)=\phi_i^*(\vecr)+\eta Y_i^*(\omega,\vecr)$ and
$\psi_i(\vecr)=\phi_i(\vecr)+\eta X_i(\vecr,\omega)$.
Exchanging the forward and backward amplitudes
in $\psi_i(\vecr)$ and $\psi'_i(\vecr)$, we may calculate 
$\delta h^\dagger(\vecr,\omega)\phi_i(\vecr)$ in the same way.

Adopting the fixed-$\omega$ local external field  
\begin{equation}
V_{\rm ext}(\vecr,\omega')= \delta_{\omega\omega'} F(\vecr) ,
\end{equation}
the transition strength can be obtained from the
calculated forward and backward amplitudes,
\begin{eqnarray}
\frac{dB(\omega; F)}{d\omega} &\equiv&
 \sum_n |\bra{n}F\ket{0}|^2 \delta(\omega-E_n) ,\\
\label{dBdw_r}
 &=& -\frac{1}{\pi} \mbox{Im} \sum_i\int d\vecr \left\{
      \phi_i^*(\vecr) F^\dagger(\vecr) X_i(\vecr,\omega)
      + Y_i^*(\vecr,\omega) F^\dagger(\vecr) \phi_i(\vecr) \right\} .
\end{eqnarray}


We apply the present method to the BKN interaction which
contains two-body (zero- and finite-range) and three-body interactions.
For this schematic interaction,
the spin-isospin degeneracy is assumed all the time and
the Coulomb potential acts on all orbitals with a charge $e/2$ \cite{BKN76}.
The HF one-body Hamiltonian in the coordinate-space representation
is given by
\begin{equation}
\label{BKN}
h[\rho]=\frac{1}{2m}\nabla^2 + \frac{3}{4}t_0\rho(\vecr)
                            + \frac{3}{16}\rho^2(\vecr)
+W_Y[\rho](\vecr) + W_C[\rho](\vecr) ,
\end{equation}
where the Yukawa potential, $W_Y$, and Coulomb potential, $W_C$,
consist of their direct terms only.
For the finite amplitude approach, it is convenient to rewrite
Eq. (\ref{BKN}) as
\begin{eqnarray}
\label{BKN2}
h[\psi'^*,\psi](\vecr)&=&\frac{1}{2m}\nabla^2 +
\frac{3}{4}t_0 \sum_{i=1}^{A/4} 4 \psi_i(\vecr) \psi_i'^*(\vecr)
+\frac{3}{16}\left\{\sum_{i=1}^{A/4} 4 \psi_i(\vecr) \psi_i'^*(\vecr)\right\}^2
\nonumber\\
&&\quad\quad + \int d\vecr' v(\vecr-\vecr')
                \sum_{i=1}^{A/4} 4\psi_i(\vecr')\psi'^*_i(\vecr') ,
\end{eqnarray}
where $v(\vecr)$ is a sum of the Yukawa and the Coulomb potential,
\begin{eqnarray}
v(\vecr)\equiv V_0 a \frac{e^{-r/a}}{r} + \frac{(e/2)^2}{|\vecr|} .
\end{eqnarray}
We adopt the parameter values from Ref.~\cite{BKN76}.

\subsection{Numerical details}

We use the three-dimensional (3D) coordinate-space representation for
solving the RPA equations.
The model space is a sphere of radius of 10 fm,
discretized in square mesh of
$\Delta x=\Delta y=\Delta z=0.8$ fm.
The number of grid points in the sphere is 8217.
The differentiation is approximated by the nine-point formula.
The frequency $\omega$ is varied from zero to 40 MeV with a spacing of
$\Delta\omega=200$ keV (201 points).
A small imaginary part is added to $\omega$:
$\omega\rightarrow \omega+i\Gamma/2$ with $\Gamma=500$ keV.
In numerical calculations,
we use real variables with double precision ($8$ bytes) and
complex variables of $8\times 2$ bytes.
In Eq.~(\ref{delta_h_3}), we choose
the parameter $\eta$ 
in $\psi_i(\vecr)=\phi_i(\vecr)+\eta X_i(\vecr)$ and
$\psi_i'^*(\vecr)=\phi_i^*(\vecr)+\eta Y_i^*(\vecr)$,
as follows:
\begin{equation}
\eta= \frac{10^{-5}}{\max\{ N(X),N(Y) \}}, \quad\quad
N(\delta\psi)\equiv\frac{1}{A}
\sqrt{\sum_i \inproduct{\delta\psi_i}{\delta\psi_i},\}} .
\end{equation}
In order to obtain the forward and backward amplitudes at a frequency $\omega$,
we adopt the Bi-CGM as an iterative solver for
Eqs. (\ref{RPA1_2}) and (\ref{RPA2_2}),
starting from the initial values of $X_i(\vecr)=Y_i^*(\vecr)=0$.
We set the convergence condition that the ratio of the remaining difference
to the r.h.s. of Eqs. (\ref{RPA1_2}) and (\ref{RPA2_2}) is less than $10^{-5}$.
The number of iteration necessary to reach the convergence
depends on the choice of the external field $V_{\rm ext}(\omega)$,
the frequency $\omega$, the smoothing parameter $\Gamma$,
and residual interactions included in the calculation.
The convergence is relatively slow for an external field
coupled to the NG modes.
A larger number of iteration is required for a larger $\omega$ value.
Typically, the calculation reaches the convergence in 10 to 100 iterations
for $\omega < 10$ MeV, but it requires more than
500 iterations for $\omega > 30$ MeV.
The number also depends on the smoothing parameter $\Gamma$.
Roughly speaking, larger number of iteration seems to be required
for smaller $\Gamma$.
If we neglect the residual Coulomb and Yukawa interactions of finite range,
the convergence becomes much faster.
We solve the differential equations to obtain the Coulomb and Yukawa
potentials using the CGM \cite{FKW78}.
\begin{equation}
\label{diff_VCY}
\nabla^2 V_C = -2\pi e^2\rho(\vecr) ,\quad\quad
\left(\nabla^2 -\frac{1}{a^2}\right) V_Y = -4\pi V_0 a \rho(\vecr) .
\end{equation}
It turns out to be important to solve these equations with high accuracy.
We set the convergence condition that the ratio of the remaining difference
to the r.h.s. of Eq. (\ref{diff_VCY}) is less than $10^{-23}$.
Since the convergence of the CGM is very fast, this is not a problem.

\subsection{Results}

In this section, we show calculated response for isoscalar (IS) modes of
compressional dipole, quadrupole and octupole for $^{20}$Ne.
The main purpose of the calculation is to test capability of
the present numerical approach,
the finite amplitude method.
The $^{20}$Ne nucleus has a prolate shape with a quadrupole deformation
$\beta\approx 0.4$ in the HF ground state.
Identifying the symmetry axis with $z$-axis, we use
external fields with a fixed frequency,
$V_{\rm ext}(\vecr,\omega)=Q_{\lambda K}(\vecr)$,
\begin{equation}
Q_{\lambda K}(\vecr)=\begin{cases}
r^3 Y_{10}(\hat\vecr), \quad r^3 Y_{11}(\hat\vecr), &
                           \text{for IS dipole, }\lambda=1 \\
r^2 Y_{20}(\hat\vecr), \quad r^2 Y_{21}(\hat\vecr), \quad r^2 Y_{22}(\hat\vecr),
                    & \text{for IS quadrupole, }\lambda=2 \\
r^3 Y_{30}(\hat\vecr), \quad r^3 Y_{31}(\hat\vecr), \quad r^3 Y_{32}(\hat\vecr),
    \quad r^3 Y_{33}(\hat\vecr), & \text{for IS octupole, }\lambda=3 .
      \end{cases}
\end{equation}
Then, the strength distribution,
\begin{equation}
\frac{dB(\omega; Q_{\lambda K})}{d\omega} =
 \sum_n |\bra{n} Q_{\lambda K}\ket{0}|^2 \delta(\omega-E_n) ,
\end{equation}
will be calculated according to Eq. (\ref{dBdw_r}).

\subsubsection{Isoscalar quadrupole response:
Accuracy of the finite amplitude method}

\begin{figure}[t]
\includegraphics[width=0.53\textwidth]{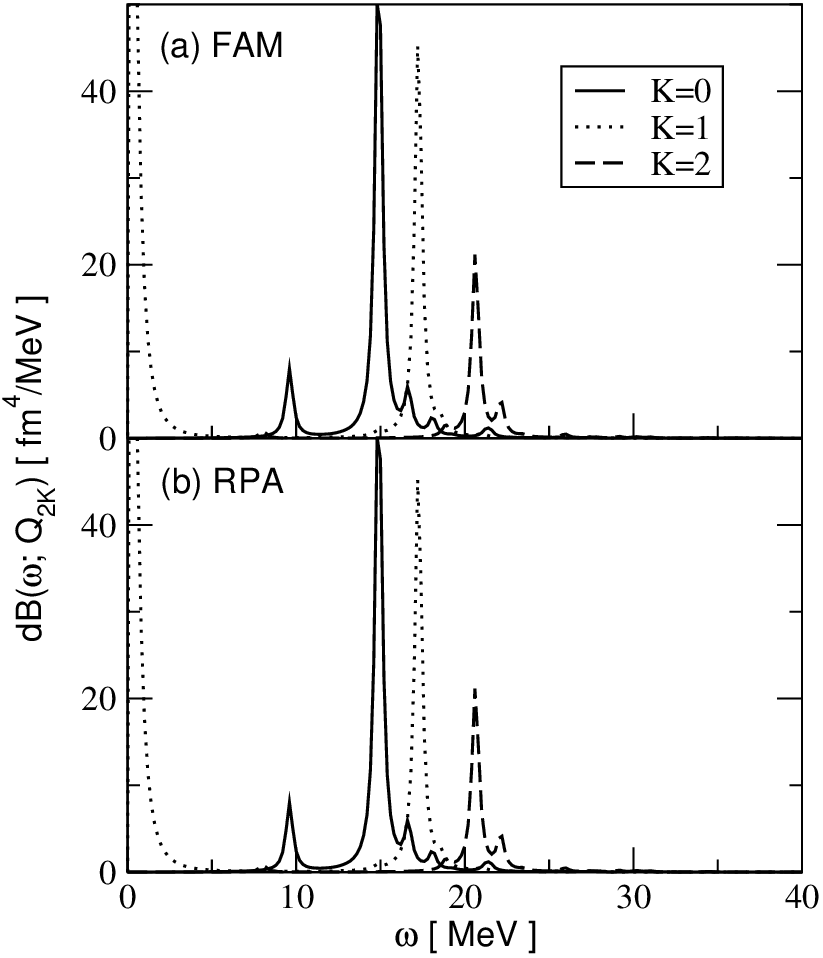}
\caption{\label{fig: quadrupole}
IS quadrupole strength distribution for $^{20}$Ne.
The solid, dotted, dashed lines indicate those with
$K=0$, 1, and 2, respectively.
Results of two kinds of calculations,
(a) the finite amplitude method (FAM) and
(b) the conventional RPA, are compared.
}
\end{figure}

In Fig.~\ref{fig: quadrupole}(a), we show results for the IS quadrupole
strength distribution.
There is a NG mode in the $K=1$ sector,
corresponding to the nuclear rotation.
This is clearly seen in the response of the $K=1$ mode,
having a large peak near $\omega=0$.
The RPA correlation brings the lowest one-particle-one-hole (1p1h)
excitation at $E_x=4.5$ MeV down to zero.
The response function for the $K=1$ mode was not obtained by
the small-amplitude TDHF method in Ref.~\cite{NY05},
because the nucleus actually rotates in real time, that violates
the small-amplitude approximation.
This is an advantage of the present method over the time-dependent
approach.
The lowest intrinsic (physical) excitation corresponds to
the $K=2$ mode at $\omega=8$ MeV which
is close to energy of the 1p1h excitation.
This suggests that the correlation effect is weak for this mode,
supported by a small $K=2$ quadrupole strength at $\omega=8$ MeV.
In contrast, the next lowest mode at $E_x=9.6$ MeV with $K=0$ is somewhat
lowered by the correlation and exhibit a larger strength.
Ref.~\cite{SONY06} shows results of configuration mixing calculation
with the BKN interaction, indicating $J^\pi=0^+$ around $E_x=7$ MeV 
and $J^\pi=2^+$ state near 8 MeV.
Large peaks at $\omega=15\sim 22$ MeV should correspond to
the IS giant quadrupole resonance.
It clearly shows deformation splitting; the $K=0$ peak at lowest, the $K=1$
in the middle, and the $K=2$ at the highest energy.

Now, let us demonstrate accuracy of
the present finite amplitude method.
In Fig.~\ref{fig: quadrupole}, results of
the conventional RPA, which explicitly estimates the residual interactions
$\delta h/\delta \rho$,
are also presented in the panel (b).
These two kinds of calculations, (a) and (b), provide identical results
in the accuracy of three to four digits.

\subsubsection{Isoscalar dipole and octupole response:
Removal of NG modes}

\begin{figure}[t]
\includegraphics[width=0.7\textwidth]{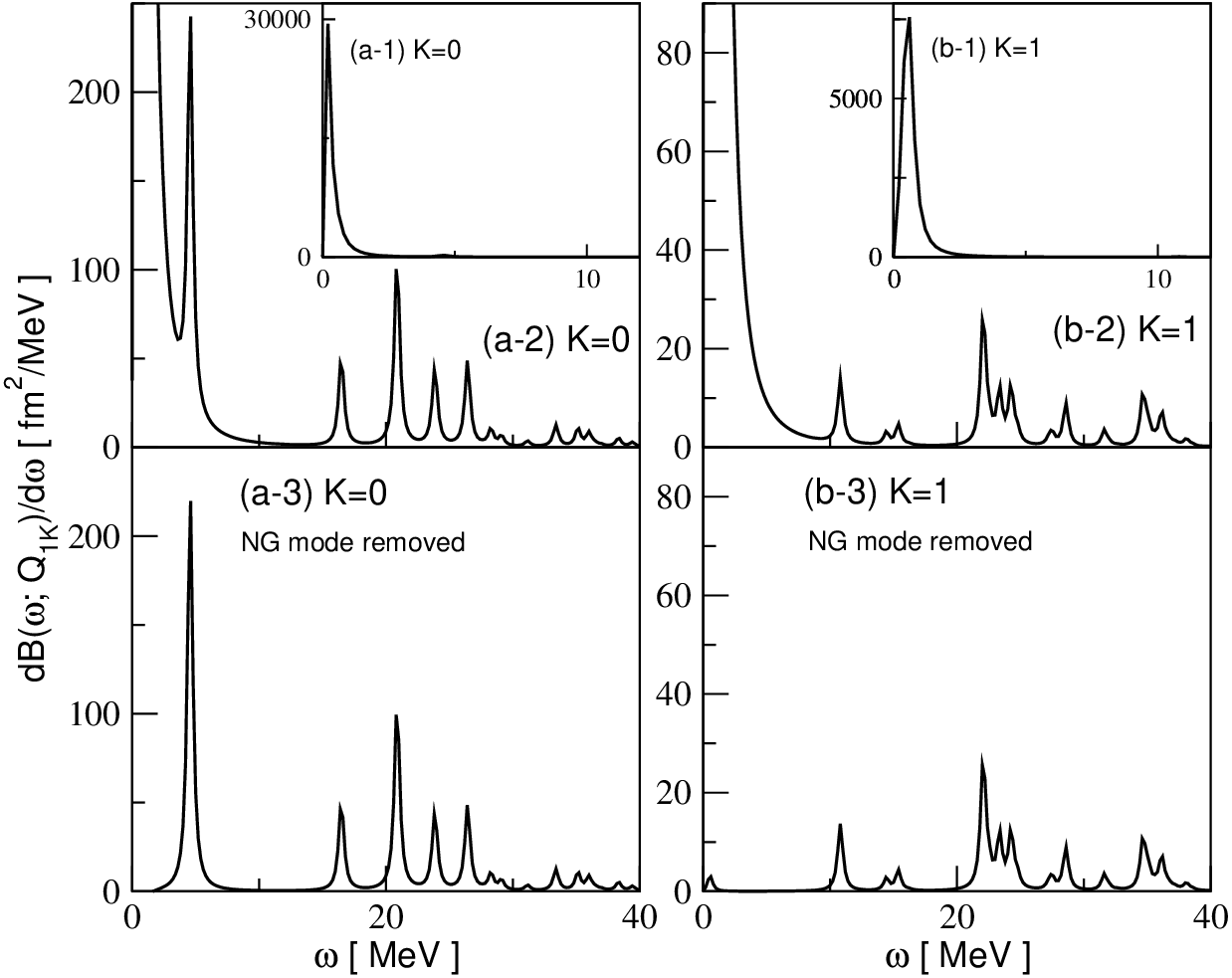}
\caption{\label{fig: dipole}
IS compressional dipole strength distribution for $^{20}$Ne.
Strengths associated with the $K=0$ modes are shown in the left (a)
and those with $K=1$ in the right (b).
The upper panels show strengths calculated with
$\delta\rho_{\rm cal}(\omega)$, while
the lower panels, (a-3) and (b-3), show those calculated with
$\delta\rho_{\rm phy}(\omega)$ in Eq. (\ref{delta_rho_phy}).
See text for details.
}
\end{figure}
\begin{figure}[ht]
\includegraphics[width=0.9\textwidth]{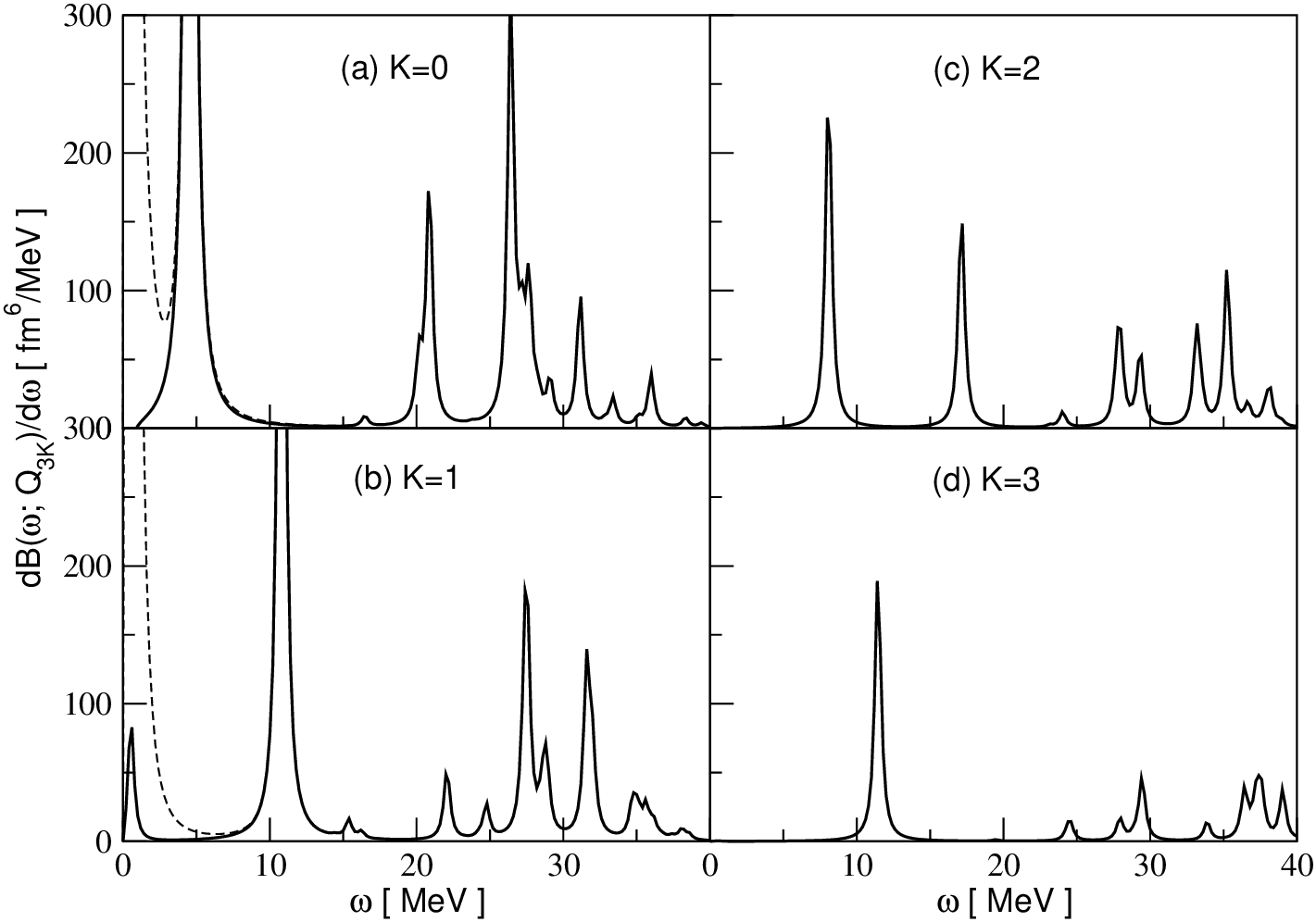}
\caption{\label{fig: octupole}
IS octupole strength distribution for $^{20}$Ne.
Strengths associated with the $K=0$, 1, 2, and 3 octupole modes are shown
in the panels (a), (b), (c), and (d), respectively.
In the panels (a) and (b), the strengths calculated with
$\delta\rho_{\rm cal}(\omega)$  are presented by dotted lines,
while those with $\delta\rho_{\rm phy}(\omega)$ are by solid lines.
}
\end{figure}

Next, we show the strength distribution for the isoscalar
compressional dipole mode.
This mode has been of significant interest because its energy is related
to the compressibility of nuclear matter, providing
information independent from the monopole resonance.
The compressional modes in spherical nuclei have been extensively studied with
the continuum RPA calculations \cite{GS81,HSZ98,SS02}.
However, these calculations are not fully self-consistent, thus,
need to remove mixture of the NG (translational) components
by modifying the dipole operator.
This produces some ambiguity in their results.
In fact, the importance of the full self-consistency
has been stressed for the compressional modes \cite{ASS03,AS04}.
So far, our understanding of the compressional dipole mode is still obscure
and further studies are needed.
In this section, we show a fully self-consistent calculations for
deformed nuclei.

In Fig.~\ref{fig: dipole},
the compressional dipole strength is shown, $K=0$ mode at the left (a) and
$K=1$ at the right (b).
There are the NG modes associated with the translational
symmetry breaking near $\omega=0$, seen in
Figs.~\ref{fig: dipole} (a-1) and (b-1).
These NG peaks are so huge that other peaks are invisible in these insets.
The vertical axis is magnified in the panels (a-2) and (b-2).
The giant resonance peaks are spread over $\omega=16\sim 30$ MeV for
$K=0$ and $20-40$ MeV for $K=1$.
There is a sharp peak at $\omega=4.5$ MeV, which
is embedded in the tail of the NG mode.
In order to estimate the strength carried by this state,
we need to separate out the contribution from the NG mode.
This is done by using the prescription described in Sec. \ref{sec: NG_mode},
adopting the center-of-mass coordinates and the total linear momenta
(3 NG modes).
Strength associated with the ``physical'' transition density
$\delta\rho_{\rm phy}(\omega)$ is shown in Fig.~\ref{fig: dipole} (a-3)
and (b-3).
The large strength of translational modes near $\omega=0$
is properly removed.
The other physical peaks with finite $\omega$ are unchanged, which indicates
that there is very little mixture of the NG modes because
our calculation is fully self-consistent.
Now, we may identify the $K=0$ peak at $\omega=4.5$ MeV as an isolated
peak.

Finally, we show IS octupole strength distribution
with $K=0$, 1, 2, and 3 in Fig.~\ref{fig: octupole}.
The lowest octupole state is at $\omega=4.5$ MeV with $K=0$, and the second
lowest is at $\omega=8.1$ MeV with $K=2$.
These results are similar to that of the variation-after-parity-projection
calculation \cite{TYI96} and that of the configuration-mixing
calculation \cite{SONY06}.
Experimentally, the band head of the $K=2$ band ($J^\pi=2^-$)
is observed at $E_x=5.0$ MeV
and that of $K=0$ ($J^\pi=1^-$) is at $E_x=5.8$ MeV.
The BKN interaction, that does not contain the spin-orbit force,
is able to reproduce the $K=0$ state in a reasonable accuracy,
however, fails to provide a quantitative description for the $K=2$ state.
This suggests that the spin-orbit force does not play an important role
for the $K=0$ state.
In fact, the parity-projected HF calculation with the Skyrme interaction
has confirmed very small contribution of the spin-orbit force in this
$K^\pi=0^-$ state \cite{OYN04}.

Since the nucleus is deformed, the dipole modes are coupled to
the octupole modes.
We may identify peaks at the same positions
in Figs. \ref{fig: dipole} and \ref{fig: octupole}
for the $K=0$ and $K=1$.
We see a small spurious $K=1$ peak near $\omega=0$,
even after removing the NG components
(solid line in Fig. \ref{fig: octupole}(b)).
However, the peak height of the NG mode is about 5,000 fm$^6/$MeV.
Thus, more than 98 \% of the NG strength is actually removed.
We can say that
the method in Sec.~\ref{sec: NG_mode} also works for octupole modes.

\section{Conclusion and discussion}
\label{sec: conclusion}

We have presented a new numerical approach to the RPA calculation,
``finite amplitude method''.
The finite amplitude method does not require
complicated programming for complex residual interactions.
Instead, it resorts to the numerical derivation of the
residual interaction (induced field),
$\delta h/\delta\rho \cdot \delta\rho$.
The most advantageous feature of the present method is
its feasibility of programming a computer code.
The RPA calculation can be accomplished with a minor extension of the
static HF computer code,
to construct the HF Hamiltonian
with independent bra and ket single-particle states.

Here, we would like to make a remark on the meaning of different
bra and ket states.
This does not mean matrix elements between
different Slater determinants which are rather complicated.
These ``off-diagonal'' elements are necessary for
configuration-mixing calculations, such as the generator-coordinate method.
The finite amplitude method does not require these.
All we need is ``diagonal'' matrix elements of a certain one-body operator,
$\hat{o}$, in the linear order with respect to variation of the
single-particle states,
\begin{equation}
\sum_{i=1}^A \bra{\psi_i} \hat{o} \ket{\psi_i}
 \approx \sum_{i=1}^A \bra{\phi_i} \hat{o} \ket{\phi_i}
 + \sum_{i=1}^A \bra{\delta\psi_i} \hat{o} \ket{\phi_i}
 + \sum_{i=1}^A \bra{\phi_i} \hat{o} \ket{\delta\psi_i} .
\end{equation}
In order to calculate the second and the third terms in the r.h.s.
separately,
we need to input independent bra and ket single-particle states.
This can be achieved by a minor extension of the static HF code.

The method has been applied to calculations of
the isoscalar dipole, quadrupole, and octupole response functions.
Since the adopted interaction is rather schematic, we do not discuss here
calculated properties of these modes.
Instead, we would like to emphasize characteristic features
of the finite amplitude method.
First of all, the transition density coupled to the NG modes
can be calculated without any special treatment.
As is seen in the quadrupole ($K=1$) and octupole modes ($K=0$ and 1)
in Figs.~\ref{fig: quadrupole} and \ref{fig: octupole},
the NG modes appear in responses to a variety of external fields,
especially for deformed nuclei.
This causes a serious problem in the time-dependent calculation of
the small-amplitude TDHF \cite{NY05}.
If we want to remove the strength associated with the NG modes,
we can use the prescription presented in Sec.~\ref{sec: NG_mode}.
This has been demonstrated in the dipole and octupole strength distributions.
Second, since we do not calculate the residual interaction explicitly,
it is easy to carry out the fully self-consistent RPA calculation 
for realistic interactions including spin-orbit, derivative, and Coulomb terms.
The implementation of the present method does not depend on the
complexity of the interactions.
For instance, the compressional dipole mode has been a long-standing problem
in microscopic calculations \cite{Giai01}.
The problem is related to difficulties in the fully self-consistent treatment
and the coupling to the translational modes.
Our new approach may provide a tool to clarify this point.
Last but not least, the finite amplitude method is an efficient
method to calculate the strength distribution.
We may control the necessary energy resolution by the smoothing parameter
$\Gamma$.
The numerical application to the BKN functional shows that its efficiency is
next to the time-dependent method, better than the other methods including
the Green's function method \cite{NY05} and
the diagonalization method \cite{Muta02}.
The diagonalization of the RPA matrix is very efficient if we are interested
in only a few lowest states, however, it becomes more and more difficult
for higher excitation energies.

For future developments, it is interesting to combine the present method with
the absorbing-boundary-condition approach in Ref.~\cite{NY05}.
This enables us to calculate response in the continuum, overcoming
difficulties in the time-dependent method.
It is also very interesting to extend the method in the HF scheme
to the one in the Hartree-Fock-Bogoliubov framework.
In this paper, we adopt a simple interaction to check the method,
but the finite amplitude method
shows its real power for a complex density functional.
Applications of the method
to the realistic Skyrme functionals are under progress at present
and will be published in near future.

\begin{acknowledgments}
This work is supported by the Grant-in-Aid for Scientific Research
in Japan (Nos. 17540231, 18540366, 18036002), by the PACS-CS project of
Center for Computational Sciences, University of Tsukuba,  and
by the Large Scale Simulation Program No. 06-14 (FY2006)
of High Energy Accelerator Research Organization (KEK).
A part of the numerical calculations
have been also performed on the supercomputer at the Research Center for
Nuclear Study (RCNP), Osaka University, and
at YITP, Kyoto University.
\end{acknowledgments}

\bibliography{myself,nuclear_physics,chemical_physics}

\end{document}